\documentstyle[12pt,tighten,eqsecnum,epsf,aps]{revtex}

\begin{document}
\draft

\title{Coherent states for quantum systems with a trilinear \\
boson Hamiltonian}
\author{C. Brif \ \cite{email}}
\address{Department of Physics, Technion -- Israel Institute of 
Technology, Haifa 32000, Israel}
   \maketitle

	\begin{abstract}
We introduce a set of coherent states which are associated with quantum 
systems governed by a trilinear boson Hamiltonian. These states are 
produced by the action of a nonunitary displacement operator on a
reference state and can be equivalently defined by some eigenvalue
equations. The system prepared initially in the reference state will
evolve into the coherent state during the first instants of the 
interaction process. Some properties of the coherent states are 
discussed. In particular, the resolution of the identity is derived and 
the related analytic representation in the complex plane is developed. 
It is shown that this analytic representation coincides with a double 
representation based on the Glauber coherent states of the pump mode 
and on the SU(1,1) Perelomov coherent states of the signal-idler system. 
Entanglement between the field modes and photon statistics of the 
coherent states are studied. Connections between the coherent states
and the long-time evolution induced by the trilinear Hamiltonian are
considered.
	\end{abstract}


\section{Introduction}
\label{sec:intr}

Various nonlinear processes in quantum optics, such as parametric
amplification, frequency conversion, Raman and Brillouin scattering, 
and the interaction of two-level atoms with a single-mode radiation 
field, are described by a trilinear boson Hamiltonian of the form
\cite{MG67,Gr68,TC68,TW69,WB70,BP70,FG71,Lu73,SLu74,AM74,PV75,%
Pe76,Ga77,KM80,MC81,KH81,HDR82,McNG83,GH84,Tu86,Ju89,Ca89,GM91,%
Pe91,DJ92b,JD92,DJB93,HYB95,BDJ95}
\begin{equation}
\hat{H} = \omega_a \hat{a}^{\dagger} \hat{a} + 
\omega_b \hat{b}^{\dagger} \hat{b} + \omega_c \hat{c}^{\dagger} \hat{c} 
+ \kappa  \hat{a} \hat{b}^{\dagger} \hat{c}^{\dagger} + 
\kappa^{\ast} \hat{a}^{\dagger} \hat{b} \hat{c} .  \label{h1}
\end{equation}
Here $\hat{a}$, $\hat{b}$, $\hat{c}$ are the boson annihilation operators
of the three field modes with angular frequencies $\omega_a$, $\omega_b$,
$\omega_c$, respectively, which satisfy the energy conservation law
\begin{equation}
\omega_a = \omega_b + \omega_c .  
\end{equation}

The concept of coherent states has been introduced by Glauber \cite{Gla}
for the single-mode quantized radiation field whose dynamical symmetry 
group is the Weyl-Heisenberg group \cite{Weyl}. Perelomov \cite{Per} and
Gilmore \cite{Gil,Gil:rev} have generalized the concept of coherent 
states to quantum systems whose dynamical symmetry group is an arbitrary 
Lie group. 

In the present paper we introduce coherent states for quantum systems
whose dynamics is governed by the trilinear boson Hamiltonian (\ref{h1}).
This Hamiltonian is a linear combination of operators that do not form a
finite-dimensional Lie algebra, and hence a direct application of the
Perelomov formalism becomes impossible in this case. Therefore, we 
define the coherent states as eigenstates of some non-Hermitian operators 
or, equivalently, as states produced by the action of a nonunitary 
displacement operator on a reference state. We show that during the 
first instants of the time evolution (i.e., in the short-time 
approximation) this reference state is evolved into the coherent state.
Moreover, the coherent states have an important property of resolution 
of the identity. Thus we are able to develop an analytic representation 
in the complex plane. Some quantum statistical properties of the 
coherent states are also discussed. In particular, we study photon 
statistics of the pump mode and the entanglement between the pump and 
the signal-idler subsystems. We also consider connections between the
coherent states and the long-time dynamics governed by the trilinear 
boson Hamiltonian.

\section{Physical models}
\label{sec:mod}

The dynamics induced by the Hamiltonian (\ref{h1}) can be better 
understood by revealing its symmetries. It can be easily shown 
that the operators 
\begin{mathletters}
\begin{eqnarray}
& & \hat{{\cal S}}_{ab} = \hat{a}^{\dagger}\hat{a} 
+ \hat{b}^{\dagger}\hat{b} , \\
& & \hat{{\cal S}}_{ac} = \hat{a}^{\dagger}\hat{a} 
+ \hat{c}^{\dagger}\hat{c} , \\
& & \hat{{\cal D}}_{bc} = \hat{b}^{\dagger}\hat{b} 
- \hat{c}^{\dagger}\hat{c}
\end{eqnarray}
\end{mathletters}
(known in literature as Manley-Rowe invariants) are integrals of 
motion. They manifest underlying SU(2) and SU(1,1) symmetries. 
Only two of these invariants are linearly independent:
\[
\hat{{\cal S}}_{ab} - \hat{{\cal S}}_{ac} = \hat{{\cal D}}_{bc} .
\]
Therefore, a state of the system can be characterized by two quantum
numbers. The Hamiltonian (\ref{h1}) can be written in the form
\begin{equation}
\hat{H} = \hat{H}_0 + \hat{H}_{{\rm int}} 
\end{equation}
where
\begin{mathletters}
\begin{eqnarray}
& & \hat{H}_0 = \omega_a \hat{a}^{\dagger}\hat{a} + 
\omega_b \hat{b}^{\dagger}\hat{b} + 
\omega_c \hat{c}^{\dagger}\hat{c} , \\
& & \hat{H}_{{\rm int}} = \kappa  
\hat{a}\hat{b}^{\dagger}\hat{c}^{\dagger} 
+ \kappa^{\ast} \hat{a}^{\dagger}\hat{b}\hat{c} , 
\end{eqnarray}
\end{mathletters}
with 
\begin{equation}
[ \hat{H}_0 , \hat{H}_{{\rm int}} ] = 0 .
\end{equation}

With the choice $\omega_b = \omega_c = \omega_a /2$, the Hamiltonian 
(\ref{h1}) can be written (up to an unimportant constant) as 
\begin{equation}
\hat{H} = \omega_a ( \hat{a}^{\dagger}\hat{a} + \hat{K}_0 ) +
\kappa  \hat{a} \hat{K}_+ + \kappa^{\ast} \hat{a}^{\dagger} \hat{K}_- ,
\label{h2}
\end{equation}
where the operators 
\begin{equation}
\hat{K}_+ = \hat{b}^{\dagger}\hat{c}^{\dagger} , \mbox{\hspace{0.4cm}}
\hat{K}_- = \hat{b}\hat{c} , \mbox{\hspace{0.4cm}}
\hat{K}_0 = \mbox{$\frac{1}{2}$} ( \hat{b}^{\dagger}\hat{b} + 
\hat{c}^{\dagger} \hat{c} + 1)   \label{k1}
\end{equation}
span the SU(1,1) Lie algebra:
\begin{equation}
[ \hat{K}_0 , \hat{K}_{\pm} ] = \pm \hat{K}_{\pm} , 
\mbox{\hspace{0.4cm}}
[ \hat{K}_- , \hat{K}_+ ] = 2 \hat{K}_0 .  
\end{equation}
The form (\ref{h2}) is convenient for description of processes with 
underlying SU(1,1) symmetry, e.g., for parametric amplification where
the modes $\hat{a}$, $\hat{b}$, and $\hat{c}$ are identified as the pump,
signal, and idler modes, respectively. The signal-idler system possesses
an SU(1,1) symmetry, and the Casimir operator
\begin{equation}
\hat{K}^2 = \hat{K}_0^2 - \mbox{$\frac{1}{2}$} ( \hat{K}_+ \hat{K}_- +
\hat{K}_- \hat{K}_+ ) = k(k-1) 
\end{equation}
is an integral of motion:
\begin{equation}
\hat{K}^2 = \mbox{$\frac{1}{4}$} \hat{{\cal D}}_{bc}^{2} - 
\mbox{$\frac{1}{4}$} .
\end{equation}
Thus the Bargmann index $k$ that labels irreducible representations of
SU(1,1) is related to the photon-number difference between the signal
and idler modes:
\begin{equation}
k = \mbox{$\frac{1}{2}$} ( |\hat{{\cal D}}_{bc}| + 1 ) .  \label{k}
\end{equation}
Thus $k$ takes discrete values $k= \frac{1}{2},1,\frac{3}{2},2,\ldots$ 
that corresponds to the so-called discrete series of representations 
\cite{SU11}. The representation Hilbert space is spanned by the complete 
orthonormal basis $|k,n\rangle$ ($n=0,1,2,\ldots$),
\[
\hat{K}_0 |k,n\rangle = (k+n) |k,n\rangle .   
\]
For the two-mode realization (\ref{k1}) the states $|k,n\rangle$ can be
expressed in terms of Fock states of the modes:
\begin{equation}
|k,n\rangle = |n+2k-1\rangle_b |n\rangle_c .  \label{knstate}
\end{equation}
This form implies that there is more photons in the mode $\hat{b}$ than 
in the mode $\hat{c}$. However, since the Hamiltonian is symmetric in 
the operators $\hat{b}$ and $\hat{c}$, we may by definition refer to the 
mode with larger number of photons as the mode $\hat{b}$ and to the 
mode with smaller number of photons as the mode $\hat{c}$. Certainly,
this definition is incomplete if an initial state is given, for example, 
by a superposition $|n\rangle_b |0\rangle_c + |0\rangle_b |m\rangle_c$. 
In such a case we must use two SU(1,1) representations:
$|k,n\rangle_{+} = |n+2k-1\rangle_b |n\rangle_c$ and 
$|k,n\rangle_{-} = |n\rangle_b |n+2k-1\rangle_c$. However, for the sake
of simplicity, we will not consider such situations.

There is a number of nonlinear optical processes described by the 
Hamiltonian (\ref{h2}). For degenerate parametric amplification or 
second harmonic generation, one has
\begin{equation}
\hat{K}_+ = \mbox{$\frac{1}{2}$} \hat{b}^{\dagger 2} , 
\mbox{\hspace{0.4cm}}
\hat{K}_- = \mbox{$\frac{1}{2}$} \hat{b}^{2} , \mbox{\hspace{0.4cm}}
\hat{K}_0 = \mbox{$\frac{1}{2}$} ( \hat{b}^{\dagger} \hat{b} + 
\mbox{$\frac{1}{2}$} ) .  \label{k2}
\end{equation}
In this case $\hat{K}^2 = -3/16$, so $k=1/4$ or $3/4$. The dynamics 
induced by the corresponding Hamiltonian has been studied for both
second harmonic generation \cite{WT72,MR78} and degenerate parametric
amplification \cite{HZ84,CrB88,KFD93,HYB94,CoB95}. 
For the intensity-dependent two-mode coupling, the SU(1,1) generators 
are given by
\begin{equation}
\hat{K}_+ = \sqrt{\hat{b}^{\dagger}\hat{b}}\,\, \hat{b}^{\dagger} ,
\mbox{\hspace{0.4cm}} 
\hat{K}_- = \hat{b} \sqrt{\hat{b}^{\dagger}\hat{b}} , 
\mbox{\hspace{0.4cm}} 
\hat{K}_0 = \hat{b}^{\dagger}\hat{b} + \mbox{$\frac{1}{2}$} ,
\label{k3}
\end{equation}
which is known as the Holstein-Primakoff realization \cite{HP40,Ge83}.
In this case $k=1/2$.

The Hilbert space ${\cal H}$ of a quantum system governed by the
Hamiltonian (\ref{h2}) can be decomposed into a direct sum of 
finite-dimensional subspaces ${\cal H}_L$:
\begin{equation}
{\cal H} = \bigoplus_{L=0}^{\infty} {\cal H}_L .
\end{equation}
Each subspace ${\cal H}_L$ is spanned by the complete orthonormal set 
of $(L+1)$ states $|n\rangle |k,L-n\rangle$ $(n=0,\ldots,L)$, where 
$|n\rangle$ are the Fock states of the mode $\hat{a}$,
\[
\hat{a}^{\dagger} \hat{a} |n\rangle = n |n\rangle , 
\]
and $|k,L-n\rangle$ are the orthonormal states of SU(1,1). We have seen
that the Bargmann index $k$ is related to the integral of motion 
$\hat{{\cal D}}_{bc}$ via Eq.\ (\ref{k}). Correspondingly, we find the 
relation between the integrals of motion and the quantum number $L$:
\begin{equation}
L = \mbox{$\frac{1}{2}$} ( \hat{{\cal S}}_{ab} + \hat{{\cal S}}_{ac} 
- \hat{{\cal D}}_{bc} ) .
\end{equation}
It is clear that the operator
\begin{equation}
\hat{H}_0 = \omega_a ( \hat{a}^{\dagger} \hat{a} + \hat{K}_0 )
\end{equation}
has the property
\begin{equation}
\hat{H}_0 |n\rangle|k,L-n\rangle = \omega_a (k+L) |n\rangle|k,L-n\rangle .
\end{equation}
Since $\hat{H}_0$ commutes with
\begin{equation}
\hat{H}_{{\rm int}} = \kappa \hat{a} \hat{K}_+ + 
\kappa^{\ast} \hat{a}^{\dagger} \hat{K}_- ,
\end{equation}
the subspace ${\cal H}_L$ is invariant under the evolution induced by
the Hamiltonian (\ref{h2}).

\section{Coherent states}

The evolution operator $\hat{U}(t) = \exp(- i \hat{H} t)$ can be 
written as
\begin{equation}
\hat{U}(t) = \exp( z \hat{a} \hat{K}_+ - 
z^{\ast} \hat{a}^{\dagger} \hat{K}_- )  \exp(- i \hat{H}_0 t) ,
\label{U}
\end{equation}
where $z = - i \kappa t$. If the initial state is $|\psi_0 ;k,L\rangle$,
the action of $\exp(- i \hat{H}_0 t)$ will only multiply it by a phase
factor:
\[
\exp(- i \hat{H}_0 t) |\psi_0 ;k,L\rangle =
\exp[- i \omega_a (k+L) t ] |\psi_0 ;k,L\rangle .
\]
Therefore we can consider only the action of the first exponent in 
(\ref{U}):
\begin{equation}
|\psi_z ;k,L\rangle = \exp( z \hat{a} \hat{K}_+ - 
z^{\ast} \hat{a}^{\dagger} \hat{K}_- ) |\psi_0 ;k,L\rangle . \label{es}
\end{equation}
If the mode $\hat{a}$ is highly excited and its depletion can be 
neglected through the evolution of the system, one can use the so-called
parametric approximation \cite{BP70,KM80}. In this approximation the 
mode $\hat{a}$ is treated classically, i.e., the operators $\hat{a}$ 
and $\hat{a}^{\dagger}$ are replaced by $c$ numbers $\alpha$ and 
$\alpha^{\ast}$, respectively. In this case Eq.\ (\ref{es}) defines the
SU(1,1) Perelomov coherent states. In the context of the boson 
realizations (\ref{k1}) and (\ref{k2}), these states are identified as
the well-known two-mode and one-mode squeezed states, respectively. 
(Limitations to squeezing in degenerate parametric amplifiers due to 
quantum fluctuations in the pump have been discussed in Refs.\
\cite{HZ84,CrB88,KFD93,HYB94,CoB95}.)
The SU(1,1) coherent states have been also considered 
\cite{Ah73,Vo,Brif} in the context of the Holstein-Primakoff realization 
(\ref{k3}).

In what follows we will consider the general (pure quantum) case,
when the parametric approximation is not valid. However, it is 
customary to use the short-time approximation 
\cite{AM74,Pe76,HDR82,Pe91,DJB93,BDJ95} in which one assumes $|z| \ll 1$.
In order to clarify the algebraic meaning of this approximation, we 
use the Baker-Campbell-Hausdorff theorem expressed as the Zassenhaus 
formula \cite{Ma54}:
\begin{equation}
e^{\hat{X}+\hat{Y}} = e^{\hat{X}} e^{\hat{Y}} e^{\hat{C}_2} 
e^{\hat{C}_3} \cdots e^{\hat{C}_j} \cdots . \label{zass}
\end{equation}
Here $\hat{C}_j$ is a polynomial of $j$th degree in $\hat{X}$ and 
$\hat{Y}$ with rational coefficients. For example, 
\begin{eqnarray}
& & \hat{C}_2 = -\mbox{$\frac{1}{2}$} [\hat{X},\hat{Y}] , \\
& & \hat{C}_3 = -\mbox{$\frac{1}{3}$} [[\hat{X},\hat{Y}],\hat{Y}] 
-\mbox{$\frac{1}{6}$} [[\hat{X},\hat{Y}],\hat{X}] .
\end{eqnarray}
For the evolution operator of Eq.\ (\ref{es}), we identify
\begin{equation}
\hat{X} = z \hat{a} \hat{K}_+  , \mbox{\hspace{0.4cm}}
\hat{Y} = - z^{\ast} \hat{a}^{\dagger} \hat{K}_- .
\end{equation}
Therefore, $\hat{C}_j$ contains $|z|^j$. For example,
\[
\hat{C}_2 = \frac{|z|^2}{2}
[ \hat{a} \hat{K}_+ , \hat{a}^{\dagger} \hat{K}_- ] =
\frac{|z|^2}{2} [ \hat{K}_- \hat{K}_+ - 
2 \hat{a} \hat{a}^{\dagger} \hat{K}_0 ] .
\]
In the short-time approximation one truncates in the second order of
$|z|$, i.e., neglects all the terms that contain $|z|^j$ with $j>2$.
In this approximation we obtain
\begin{equation}
|\psi_z ;k,L\rangle \approx \left[ 1 + ( z \hat{a} \hat{K}_+
- z^{\ast} \hat{a}^{\dagger} \hat{K}_- ) + \mbox{$\frac{1}{2}$}
( z \hat{a} \hat{K}_+ - z^{\ast} \hat{a}^{\dagger} \hat{K}_- )^2
\right] |\psi_0 ;k,L\rangle . \label{es_ap}
\end{equation}

Let us consider the initial state of the form 
\begin{equation}
|\psi_0 ;k,L\rangle = |L\rangle |k,0\rangle . \label{is}
\end{equation}
Then Eq.\ (\ref{es_ap}) gives
\begin{eqnarray}
|\psi_z ;k,L\rangle & \approx & ( 1 - k L |z|^2 ) |L\rangle |k,0\rangle
+ z \sqrt{2 k L} |L-1\rangle |k,1\rangle \nonumber \\
& & + z^2 \sqrt{(2k^2 + k)(L^2 - L)} |L-2\rangle |k,2\rangle .
\label{es_ef}
\end{eqnarray}
The initial state (\ref{is}) is an extreme state in the subspace 
${\cal H}_L$ and hence it has some special properties:
\begin{equation}
e^{\hat{C}_2} |L\rangle |k,0\rangle \approx 
\left( 1 + \mbox{$\frac{1}{2}$} |z|^2 
[ \hat{a} \hat{K}_+ , \hat{a}^{\dagger} \hat{K}_- ] \right)
|L\rangle |k,0\rangle \nonumber 
= ( 1 - k L |z|^2 ) |L\rangle |k,0\rangle ,
\end{equation}
\begin{equation}
\exp( - z^{\ast} \hat{a}^{\dagger} \hat{K}_- ) |L\rangle |k,0\rangle
= |L\rangle |k,0\rangle .
\end{equation}
Therefore, we find
\begin{equation}
|\psi_z ;k,L\rangle \approx ( 1 - k L |z|^2 ) \exp(z \hat{a} \hat{K}_+)
|L\rangle |k,0\rangle ,  \label{es_lf}
\end{equation}
where the exponential should be expanded up to the second order, which
immediately leads to Eq.\ (\ref{es_ef}).

Now we introduce the coherent states for the quantum systems governed
by the Hamiltonian (\ref{h2}). We define
\begin{equation}
|z;k,L\rangle \equiv e^{\hat{X}} e^{\hat{Y}} e^{\hat{C}_2} 
|L\rangle |k,0\rangle 
= {\cal N}^{-1/2} \exp(z \hat{a} \hat{K}_+)
|L\rangle |k,0\rangle ,  \label{cs}
\end{equation}
where ${\cal N}$ is the normalization factor determined by
\[
\exp\left( \mbox{$\frac{1}{2}$} |z|^2 
[ \hat{a} \hat{K}_+ , \hat{a}^{\dagger} \hat{K}_- ] \right)
|L\rangle |k,0\rangle = {\cal N}^{-1/2} |L\rangle |k,0\rangle .
\]
Comparing Eqs.\ (\ref{cs}) and (\ref{es_lf}), we see that in the 
short-time approximation the coherent states $|z;k,L\rangle$ coincide 
with the states $|\psi_z ;k,L\rangle$ obtained by the action of
the evolution operator on the initial state $|L\rangle |k,0\rangle$.
Expanding the exponential in (\ref{cs}), we obtain
\begin{equation}
|z;k,L\rangle = \frac{1}{\sqrt{{\cal N}}} \sum_{n=0}^{L} 
\tilde{d}_{n}(k,L) z^n |L-n\rangle |k,n\rangle ,     \label{cs_z}
\end{equation}
\begin{equation}
\tilde{d}_{n}(k,L) \equiv \left[ \frac{L! \Gamma(2k+n) }{ 
n! (L-n)! \Gamma(2k) } \right]^{1/2} .
\end{equation}
Sometimes it is convenient to use $y = 1/z$. Then we write the coherent 
states in the form
\begin{equation}
|y;k,L\rangle = \frac{y^{-L}}{\sqrt{{\cal N}}} \sum_{n=0}^{L} 
d_{n}(k,L) y^n |n\rangle |k,L-n\rangle ,     \label{cs_y}
\end{equation}
\begin{equation}
d_{n}(k,L) \equiv \tilde{d}_{L-n}(k,L) = \left[ \frac{L! \Gamma(2k+L-n) 
}{ n! (L-n)! \Gamma(2k) } \right]^{1/2} .
\end{equation}
The normalization factor is given by
\[
{\cal N}(|y|^2;k,L) = \frac{ \Gamma(2k+L) }{ |y|^{2L} \Gamma(2k) }\, 
\Phi(-L;1-L-2k;|y|^2) , 
\]
where $\Phi(a;c;x)$ is the confluent hypergeometric function (the
Kummer function) \cite{Erd}. The scalar product of two coherent
states is
\[
\langle y_1 ;k,L | y_2 ;k,L \rangle = \frac{ 
{\cal N}( y_1^{\ast} y_2 ;k,L) }{ \left[ {\cal N}(| y_1 |^2 ;k,L)
{\cal N}(| y_2 |^2 ;k,L) \right]^{1/2} } . 
\]

Equation (\ref{cs}) shows that the coherent states $|z;k,L\rangle$ 
are produced by the action of the (nonunitary) displacement operator
$\exp( z \hat{a} \hat{K}_+ )$ on the reference state $|L\rangle 
|k,0\rangle$. They can be equivalently defined by some eigenvalue 
equations. We start with eigenvalue equations satisfied by the 
reference state $|L\rangle |k,0\rangle$:
\begin{mathletters}
\begin{eqnarray}
& & \hat{a}^{\dagger} \hat{a} |L\rangle |k,0\rangle = 
L |L\rangle |k,0\rangle  , \label{eieq1a} \\
& & \hat{K}_0 |L\rangle |k,0\rangle = 
k |L\rangle |k,0\rangle  , \label{eieq1b} \\
& & \hat{K}_- |L\rangle |k,0\rangle = 0  . \label{eieq1c}
\end{eqnarray}
\end{mathletters}
Acting on both sides of these equations with $\exp(z\hat{a}\hat{K}_+ )$,
we obtain the eigenvalue equations satisfied by the coherent states:
\begin{mathletters}
\begin{eqnarray}
& & ( \hat{a}^{\dagger} \hat{a} + z \hat{a} \hat{K}_+ ) |z;k,L\rangle 
= L |z;k,L\rangle  , \label{eieq2a} \\
& & ( \hat{K}_0 - z \hat{a} \hat{K}_+ ) |z;k,L\rangle = 
k |z;k,L\rangle  , \label{eieq2b} \\
& & ( \hat{K}_-  - 2 z \hat{a} \hat{K}_0 + z^2 \hat{a}^2 \hat{K}_+ )
|z;k,L\rangle = 0  .  \label{eieq2c}
\end{eqnarray}
\end{mathletters}
Note that Eqs.\ (\ref{eieq2a}) and (\ref{eieq2b}) are connected via
the relation $\hat{a}^{\dagger} \hat{a} + \hat{K}_0 = k+L$.

It is straightforward to generalize the concept of the coherent states
to the case when the reference state of the pump mode is a superposition 
of the Fock states $|L\rangle$ with different values of $L$ and the 
reference state of the signal-idler system is a superposition of
the SU(1,1) states $|k,0\rangle$ with different values of $k$:
\[  
| \psi_0 \rangle = \sum_{L=0}^{\infty} h_L |L\rangle 
\sum_{k} g_k |k,0\rangle .
\]
We assume that $| \psi_0 \rangle$ is normalized. Then the coherent state 
is given by
\begin{equation}
|z\rangle = \frac{1}{\sqrt{{\cal N}}} \exp( z \hat{a}\hat{K}_+ ) 
| \psi_0 \rangle = \sum_{k,L} g_k h_L |z;k,L\rangle . \label{cs_sup}
\end{equation}
For example, one can consider a case when the reference state of the pump
mode is the Glauber coherent state or the squeezed state.
In what follows we will study the coherent states with specific values 
of $k$ and $L$. Properties of the coherent superposition (\ref{cs_sup})
depend on the choice of the coefficients $g_k$ and $h_L$.
Expectation values for the state $|z\rangle$ are given by weighted
sums over the corresponding expectations for the states $|z;k,L\rangle$.

\section{Resolution of the identity and analytic representations}
\label{sec:ident}

An important property of coherent states is the resolution of the 
identity. For the coherent states $|y;k,L\rangle$ we have
\begin{equation}
\int d \mu(y;k,L) |y;k,L \rangle\langle y;k,L| = \hat{1}_L , \label{id}
\end{equation}
where $\hat{1}_L$ is the projection operator on the subspace 
${\cal H}_L$:
\begin{equation}
\hat{1}_L = \sum_{n=0}^{L} |n\rangle |k,L-n \rangle \langle n| 
\langle k,L-n| .
\end{equation}
In order to find the integration measure $ d \mu(y;k,L)$ that 
realizes the resolution of the identity (\ref{id}), we substitute 
$y = \sqrt{r} e^{ i \phi}$, i.e.,
\[
r = |y|^2 , \mbox{\hspace{0.4cm}} \phi = {\rm arg}\, y ,
\]
and define
\begin{equation}
 d \mu(y;k,L) = R(|y|^2;k,L) d^2 y = \mbox{$\frac{1}{2}$} 
R(r;k,L) d r d \phi .
\end{equation}
The integration is over the whole complex plane.
Integrating over the angle $\phi$, we obtain
\begin{equation}
\sum_{n=0}^{L} \frac{ \Gamma(2k+L-n) }{n! (L-n)! } \left[ 
\int_{0}^{\infty} r^n T(r;k,L) d r \right] 
|n\rangle |k,L-n \rangle \langle n| \langle k,L-n| 
= \hat{1}_L , \nonumber
\end{equation}
where we have defined
\begin{equation}
T(r;k,L) \equiv \frac{\pi L!}{\Gamma(2k+L)} \frac{ R(r;k,L) }{ 
\Phi(-L;1-L-2k;r) } .
\end{equation}
By using the integral \cite[p. 285]{Erd}, 
\begin{equation}
\int_{0}^{\infty} r^{b-1} \Phi(a;c;-r) d r = \frac{ \Gamma(b)
\Gamma(c) \Gamma(a-b) }{ \Gamma(a) \Gamma(c-b) } ,
\end{equation}
we obtain 
\[
T(r;k,L) = \frac{ \Gamma(L+2) }{ \Gamma(2k+L+1) }\, 
\Phi(L+2;2k+L+1;-r) .
\]
This gives us the final expression for the integration measure: 
\begin{equation}
 d \mu(y;k,L) = \frac{1}{2\pi} \frac{L+1}{L+2k} \Phi(-L;1-L-2k;r)
\Phi(L+2;2k+L+1;-r) d r d \phi .
\end{equation}

The resolution of the identity is important because it allows the use
of the coherent states as a basis in the state space. Let us consider
a state $|f\rangle$ in ${\cal H}_L$:
\begin{equation}
|f\rangle = \sum_{n=0}^{L} f_{n} |n\rangle |k,L-n\rangle . \label{fst}
\end{equation}
We introduce the analytic function
\begin{equation}
Y_{f}(y;k,L) =  {\cal N}^{1/2} y^L \langle f |y;k,L \rangle 
= \sum_{n=0}^{L} d_{n}(k,L) f_{n}^{\ast} y^n , \label{y_an}
\end{equation}
that determines the representation in the coherent-state basis:
\begin{equation}
|f\rangle = \int d \mu(y;k,L) \frac{ Y_{f}^{\ast}(y;k,L) }{ 
\sqrt{{\cal N}} (y^{\ast})^{L} } |y;k,L \rangle ,
\end{equation}
and the scalar product of two states is
\begin{equation}
\langle g|f \rangle = \int d \mu(y;k,L) \frac{ Y_{g}(y;k,L) 
Y_{f}^{\ast}(y;k,L) }{ {\cal N} |y|^{2L} } .
\end{equation}
We also find the reproducing kernel of this representation:
\begin{equation}
Y_{f}(y;k,L) = \int d \mu(x;k,L) \frac{ y^L {\cal N}(x^{\ast}y;k,L) 
}{ x^L {\cal N}(|x|^2 ;k,L) } Y_{f}(x;k,L) .
\end{equation}
The operators $\hat{a}\hat{K}_+$, $\hat{a}^{\dagger}\hat{K}_-$, and
$\hat{K}_0$ act in the Hilbert space of analytic functions 
$Y(y;k,L)$ as linear differential operators:
\begin{mathletters}
\begin{eqnarray}
& & \hat{a}\hat{K}_+ = -y \frac{ d^2 }{ d y^2 } + 
(L+2k-1) \frac{ d }{ d y } ,  \\
& & \hat{a}^{\dagger}\hat{K}_- = - y^2 \frac{ d }{ d y } + L y , \\
& & \hat{K}_0 = - y \frac{ d }{ d y } + k + L ,
\end{eqnarray}
\end{mathletters}
and the expression for $\hat{a}^{\dagger}\hat{a}$ follows immediately
from the relation $\hat{a}^{\dagger}\hat{a} + \hat{K}_0 = k+L$.

The coherent state $| y_0 ;k,L \rangle$ is represented by the 
analytic function
\begin{equation}
Y_{0}(y;k,L) = \frac{ y^L {\cal N}(y_{0}^{\ast}y;k,L) 
}{ \sqrt { {\cal N}(| y_0 |^2 ;k,L) } } .
\end{equation}
By using Eq.\ (\ref{eieq2a}), we find that this function satisfies the
second-order differential equation
\begin{equation}
y \frac{ d^2 Y_0 }{ d y^2 } - [ y_{0}^{\ast} y + (L+2k-1)] 
\frac{ d Y_0 }{ d y } + L y_{0}^{\ast} Y_0 = 0 .   \label{deq0}
\end{equation}
Eigenstates of the Hamiltonian (\ref{h2}) are determined by 
\begin{eqnarray}
& & \hat{H}_{0} |\nu,k,L\rangle = \omega_a (k+L) |\nu,k,L\rangle , \\
& & \hat{H}_{{\rm int}} |\nu,k,L\rangle = \nu |\nu,k,L\rangle .  
\label{seq}
\end{eqnarray}
The corresponding analytic function
\begin{equation}
Y_{\nu}(y;k,L) = {\cal N}^{1/2} y^L \langle \nu,k,L |y;k,L \rangle 
\end{equation}
satisfies the second-order differential equation
\begin{equation}
y \frac{ d^2 Y_{\nu} }{ d y^2 } + [\gamma y^2 - (L+2k-1)] 
\frac{ d Y_{\nu} }{ d y } + (\bar{\nu} - \gamma L y) Y_{\nu} = 0 ,
\label{deqnu}
\end{equation}
where
\begin{equation}
\gamma \equiv \kappa/\kappa^{\ast} , \mbox{\hspace{0.4cm}}
\bar{\nu} \equiv \nu/\kappa^{\ast} .
\end{equation}

The above results for the analytic representation in $y$ plane can
be easily transformed for the  analytic representation in $z$ plane.
We just use the expression (\ref{cs_z}) for the coherent states and 
obtain
\begin{equation}
Z_{f}(z;k,L) = {\cal N}^{1/2} \langle f|z;k,L \rangle 
= \sum_{n=0}^{L} \tilde{d}_{n}(k,L) f_{L-n}^{\ast} z^n .  
\end{equation}
It is easy to find the following relation:
\begin{equation}
Z_{f}(z;k,L) = z^L Y_{f}(y=1/z;k,L) .
\end{equation}
The corresponding differential operators are
\begin{mathletters}
\begin{eqnarray}
& & \hat{a}\hat{K}_+ = -z^3 \frac{ d^2 }{ d z^2 } + 
(L-2k-1) z^2 \frac{ d }{ d z } + 2 k L z ,  \\
& & \hat{a}^{\dagger}\hat{K}_- = \frac{ d }{ d z } , \\
& & \hat{K}_0 = z \frac{ d }{ d z } + k .
\end{eqnarray}
\end{mathletters}

It is interesting to study how the coherent states $|y;k,L\rangle$ of 
the whole three-mode system are related to the Glauber coherent states 
of the pump and to the SU(1,1) Perelomov coherent states of the 
signal-idler system. We consider this relation by using the 
corresponding analytic representations. The Glauber coherent states 
are given by \cite{Gla}
\begin{equation}
|\alpha\rangle = e^{-|\alpha|^2 /2} \sum_{n=0}^{\infty} \frac{ \alpha^n
}{ \sqrt{n!} } |n\rangle ,
\end{equation}
and the corresponding analytic representation in $\alpha$ plane 
(called the Bargmann representation) is well known 
\cite{Fo28,Ba61,Se62}. The SU(1,1) Perelomov coherent states can
be expressed as \cite{Per}
\begin{equation}
|k,\zeta\rangle = (1-|\zeta|^2)^k \sum_{n=0}^{\infty} \left[
\frac{ \Gamma(n+2k) }{ n! \Gamma(2k) } \right]^{1/2} \zeta^n 
|k,n\rangle ,
\end{equation}
where $|\zeta| < 1$. The corresponding analytic representation is 
referred to as the representation in the unit disk. We combine the
Bargmann representation and the representation in the unit disk in
order to obtain a double representation for the state 
$|f\rangle \in {\cal H}_L$:
\begin{equation}
D_{f}(\alpha,\zeta) = e^{|\alpha|^2 /2} (1-|\zeta|^2)^{-k}
\langle f| ( |\alpha\rangle |k,\zeta\rangle )   
= \sum_{n=0}^{L} \left[ \frac{ \Gamma(2k+L-n) }{ 
n! (L-n)! \Gamma(2k) } \right]^{1/2} f_{n}^{\ast} \alpha^n \zeta^{L-n} .
\end{equation}
The differential operators of the double representation are given by
\begin{mathletters}
\begin{eqnarray}
& & \hat{a}\hat{K}_+ = \zeta^2 \frac{ \partial^2 }{ \partial\zeta 
\partial\alpha } + 2k \zeta \frac{ \partial }{ \partial\alpha  } ,  \\
& & \hat{a}^{\dagger}\hat{K}_- = \alpha \frac{ \partial }{ 
\partial\zeta } , \\
& & \hat{K}_0 = \zeta \frac{ \partial }{ \partial\zeta } + k .
\end{eqnarray}
\end{mathletters}
It is easy to find the following relation:
\begin{equation}
D_{f}(\alpha,\zeta) = \frac{ \zeta^L }{ \sqrt{L!} }
Y_{f}(y=\alpha/\zeta) = \frac{ \alpha^L }{ \sqrt{L!} } 
Z_{f}(z=\zeta/\alpha) .
\end{equation}
This relation shows that the analytic representation in the 
coherent-state basis $|y;k,L\rangle$ is equivalent to the double 
representation based on the Glauber coherent states of the pump and 
on the SU(1,1) Perelomov coherent states of the signal-idler system.

\section{Mode entanglement and quantum statistics of the 
coherent states}
\label{sec:stat}

The quantum nature of the dynamics induced by the trilinear boson 
Hamiltonian leads to strong entanglement between the pump and the 
signal-idler subsystems \cite{DJB93}. We study here the entanglement
between the field modes when the whole three-mode system is in the 
coherent state $|y;k,L\rangle$. This is a pure state and therefore
the total entropy is zero. However, quantum correlations between the
modes lead to the increase of the marginal entropy of each mode.
The von Neumann quantum entropy $S_i$ of the mode $i$ ($i = a,b,c$)
is defined by \cite{We78}
\begin{equation}
S_i = -{\rm Tr}_i ( \hat{\rho}_i \ln \hat{\rho}_i ) ,
\end{equation}
where $\hat{\rho}_i$ is the reduced density operator of the $i$ mode.
In particular, the reduced density operator $\hat{\rho}_a$ of the
pump mode is 
\begin{equation}
\hat{\rho}_a = {\rm Tr}_{bc} ( \hat{\rho} ) ,
\end{equation}
where $\hat{\rho}$ is the density operator of the whole system, and
${\rm Tr}_{bc}$ denotes tracing over the signal and idler variables.
For the coherent state $|y;k,L\rangle$, we find
\begin{equation}
\hat{\rho}_a = \frac{1}{|y|^{2L} {\cal N}} \sum_{n=0}^{L} 
[d_{n}(k,L)]^2 |y|^{2n} |n \rangle\langle n| .
\end{equation}

If $S_x$ and $S_y$ are the marginal entropies of two subsystems which
compose the whole system with the total entropy $S$, the Araki-Lieb
theorem \cite{AL70} can be expressed as
\begin{equation}
| S_x - S_y | \leq S \leq S_x + S_y .  
\end{equation}
In our case $S=0$, and hence the marginal entropies of the pump and the 
signal-idler subsystems are equal:
\begin{equation}
S_{a} = S_{bc} .
\end{equation}
Also, it follows from the Araki-Lieb theorem that 
\begin{equation}
| S_b - S_c | \leq S_{bc} = S_a \leq S_b + S_c .
\end{equation}
A quantitative measure of the entanglement between two subsystems is
the index of correlation \cite{BP89}:
\begin{equation}
I_{x\mbox{-}y} = S_x + S_y - S_{xy} .
\end{equation}
In our case the index of correlation $I_{a\mbox{-}bc}$ between the pump 
and the signal-idler subsystems is equal to twice the marginal entropy 
of the pump mode: $I_{a\mbox{-}bc} = 2 S_a$. 

Instead of evaluating entropies it is convenient to evaluate the purity 
parameters of the modes:
\begin{equation}
S^{{\rm pur}}_i = 1 - {\rm Tr}_i ( \hat{\rho}_{i}^{2} ) .
\end{equation}
It can be shown that the purity parameter $S^{{\rm pur}}_i$ represents 
a lower bound for the corresponding entropy $S_i$, i.e., 
$S^{{\rm pur}}_i \leq S_i$. 
Moreover, in the present case the purity parameters satisfy relations 
valid for the entropy, such as the Araki-Lieb theorem. In particular, 
this gives
\begin{equation}
S^{{\rm pur}} = 0 \leq S^{{\rm pur}}_a = S^{{\rm pur}}_{bc} .
\end{equation}
For the coherent state $|y;k,L\rangle$, we obtain
\begin{equation}
S^{{\rm pur}}_a = 1 - \frac{1}{ |y|^{4L} {\cal N}^2 } \sum_{n=0}^{L} 
[d_{n}(k,L)]^4 |y|^{4n} . \label{pur}
\end{equation}
This quantity is presented in Fig.\ \ref{fig:pur} versus
$|z|=1/|y|$ for $k=1$ and various values of $L$. As $|z|$ increases, 
$S^{{\rm pur}}_a$ rapidly grows, reaches a maximum, and then decreases.
The greater is the initial number $L$ of the pump photons, the
higher is the maximum of $S^{{\rm pur}}_a$, i.e., the stronger is the 
entanglement. In the limit $|z| \ll 1$, the purity parameter increases
as
\begin{equation}
S^{{\rm pur}}_a \approx 4 k L |z|^2 + O(|z|^4) ,
\end{equation}
while in the limit $|z| \gg 1$, it decreases as
\begin{equation}
S^{{\rm pur}}_a \approx \frac{2L}{2k+L-1} |z|^{-2} + O(|z|^{-4}) 
\label{pplim} .
\end{equation}
We see that for $|z| \gg 1$ the entanglement between the pump and the
signal-idler subsystems is very weak. This result is expected due to
the fact that the coherent states are obtained by the repeated 
application of the annihilation operator $\hat{a}$ with no creation
operator $\hat{a}^{\dagger}$ being applied. Therefore, as $|z|$ becomes
large, the pump mode tends toward the vacuum and is then more pure and
more uncorrelated to the signal and idler modes.

We also consider photon statistics of the pump mode 
$\hat{a}$ for the whole system in the coherent state 
$|y;k,L\rangle$. The photon-number distribution is given by
\begin{equation}
P_{a}(n) = \langle n| \hat{\rho}_a |n \rangle = 
\frac{ [d_{n}(k,L)]^2 }{ {\cal N} } |z|^{2(L-n)} . \label{Pn}
\end{equation}
This distribution is plotted in Fig.\ \ref{fig:pnd} versus $n$
for $k=1$, $L=10$, and various values of $|z|=1/|y|$. We see that
$P_{a}(n)$ is very sensitive to the value of $|z|$. We can further
investigate photon statistics of the pump mode by evaluating the 
expectation value and the variance of 
the number operator $\hat{N}_a = \hat{a}^{\dagger}\hat{a}$.
It is easy to find the following expressions:
\begin{eqnarray}
& & \langle \hat{N}_a \rangle = \frac{r}{{\cal A}} 
\frac{ d {\cal A} }{ d r } ,  \\
& & \langle (\Delta \hat{N}_a)^2 \rangle = \frac{r^2}{{\cal A}} 
\frac{ d^2 {\cal A} }{ d r^2 } + \frac{r}{{\cal A}} 
\frac{ d {\cal A} }{ d r } - \left( \frac{r}{{\cal A}} 
\frac{ d {\cal A} }{ d r } \right)^2 ,  
\end{eqnarray}
where $r = |y|^2$ and we have defined
\begin{equation}
{\cal A}(r;k,L) \equiv r^L {\cal N}(r;k,L) 
= \frac{\Gamma(2k+L)}{\Gamma(2k)} \Phi(-L;1-L-2k;r) . 
\end{equation}
We define
\begin{equation}
\Omega(r;k,L) \equiv \frac{ d }{ d r } \ln \left[
\Phi(-L;1-L-2k;r) \right]
= \frac{L}{L+2k-1} \frac{\Phi(1-L;2-L-2k;r)}{\Phi(-L;1-L-2k;r)} .
\end{equation}
Then the expectation value and the variance of $\hat{N}_a$ can be 
written as
\begin{eqnarray}
& & \langle \hat{N}_a \rangle =  r \Omega ,  \label{Nm}  \\
& & \langle (\Delta \hat{N}_a)^2 \rangle = [r^2 + (2k+L)r] \Omega 
- r^2 \Omega^2 -L r .  \label{Nv}
\end{eqnarray}
Using the integrals of motion, we can also infer the information
about the signal-idler system:
\begin{equation}
\langle \hat{K}_0 \rangle = k + L - \langle \hat{N}_a \rangle ,
\mbox{\hspace{0.4cm}} \langle (\Delta \hat{K}_0)^2 \rangle =
\langle (\Delta \hat{N}_a)^2 \rangle .
\end{equation}

Numerical results are presented in Fig.\ \ref{fig:nn} where 
$\langle \hat{N}_a \rangle$ and $\langle (\Delta \hat{N}_a)^2 \rangle$
are plotted versus $|z| = 1/\sqrt{r}$ for $k=1$ and $L=10$.
We also consider some approximate results. In the limit $|z| \ll 1$, 
we obtain
\begin{eqnarray}
& & \langle \hat{N}_a \rangle \approx L - 2 k L |z|^2 + O(|z|^4) , \\
& & \langle (\Delta \hat{N}_a)^2 \rangle \approx 2 k L |z|^2 + O(|z|^4) ,
\end{eqnarray}
i.e., photon statistics of the pump mode is close to that of the
Fock state $|L\rangle$. However, as $|z|$ increases, photon 
statistics rapidly changes. In the limit $|z| \gg 1$, we obtain 
\begin{equation}
\langle \hat{N}_a \rangle \approx \langle (\Delta \hat{N}_a)^2 \rangle
\approx \frac{L |z|^{-2} }{L+2k-1} + O(|z|^{-4}) . \label{nlim}
\end{equation}
Therefore, for $|z| \gg 1$ photon statistics of the pump mode is 
close to be Poissonian. Because the purity parameter $S^{{\rm pur}}_a$ 
in this limit is very small [see Eq.\ (\ref{pplim})], i.e., the pump is 
almost disentangled from the signal-idler system, the state of the pump 
mode is close to a slightly excited pure state with Poissonian 
statistics. This result can be simply explained by the repeated 
application of the pump-mode annihilation operator with no creation
operator being applied. For $|z| \gg 1$ this leads to the strong 
depletion of the pump mode.

\section{Long-time evolution and fundamental limit on the energy
transfer}
\label{sec:long}

We know that the coherent states $|z;k,L\rangle$ with $|z| \ll 1$ are 
produced from the initial state $|L\rangle|k,0\rangle$ during the first
instants of the time evolution. However, this simple physical meaning is 
not valid in general for the coherent states with large values of $|z|$. 
Nevertheless, it should be interesting to study the connection between
these states and the long-time evolution induced by the trilinear
Hamiltonian.

We consider the dynamics of the initial state $|L\rangle|k,0\rangle$.
For general $L$ the eigenvalue equation (\ref{seq}) can be solved 
numerically by a diagonalization of the interaction Hamiltonian.
For $L=1$ and $L=2$ this can be done analytically. By studying these 
simple situations, we can understand general features of the long-time 
dynamics. For $L=1$ we obtain 
\begin{equation}
|\psi_t ;k,1\rangle = \exp(- i H_{{\rm int}} t) |1\rangle|k,0\rangle
= \cos(\lambda_1 t) |1\rangle|k,0\rangle + 
\sin(\lambda_1 t) |0\rangle|k,1\rangle ,  \label{psi_tk1}
\end{equation}
where $\kappa = i |\kappa|$, $\lambda_1 = \sqrt{2k}|\kappa|$.
We see that the evolution is periodical. Therefore, it is sufficient to
consider only the time range $0 \leq \lambda_1 t \leq 2\pi$.
The coherent state for $L=1$ is 
\begin{equation}
|z;k,1\rangle = \frac{ |1\rangle|k,0\rangle +
\sqrt{2k} z |0\rangle|k,1\rangle }{ \sqrt{1+2k|z|^2} } . 
\end{equation}
It is easy to check that this state can be written 
(up to an unimportant phase factor) in the form (\ref{psi_tk1})
with real $z=(2k)^{-1/2} \tan(\lambda_1 t)$. Thus the coherent state
$|z;k,1\rangle$ exactly describes the dynamics induced by the trilinear 
Hamiltonian. The first instants of the time evolution ($\lambda_1 t \ll 
1$) and the moments with $\lambda_1 t$ close to $\pi$ and $2\pi$ are 
described by the coherent states with $|z| \ll 1$, while at the moments 
with $\lambda_1 t$ close to $\pi/2$ and $3\pi/2$ the system evolves
into the coherent state with $|z| \gg 1$.

For $L=2$ the situation is more complicated. In this case we obtain
\begin{eqnarray}
|\psi_t ;k,2\rangle & = & \exp(- i H_{{\rm int}} t) |2\rangle|k,0\rangle
= (4k+1)^{-1} \Bigl\{ \sqrt{8k(2k+1)} 
\sin^2(\lambda_2 t) |0\rangle|k,2\rangle \nonumber \\
& & + \sqrt{2k(4k+1)} \sin(2\lambda_2 t) |1\rangle|k,1\rangle
+ [1+4k\cos^2(\lambda_2 t)] |2\rangle|k,0\rangle \Bigr\}  ,   
\label{psi_tk2}
\end{eqnarray}
where $\lambda_2 = \frac{1}{2}\sqrt{8k+2}|\kappa|$.
The coherent state for $L=2$ is given by
\[
|z;k,2\rangle = \frac{ |2\rangle|k,0\rangle 
+ \sqrt{4k} z |1\rangle|k,1\rangle + 
\sqrt{2k(2k+1)} z^2 |0\rangle|k,2\rangle }{
\sqrt{1 + 4k|z|^2 + 2k(2k+1)|z|^4} } .  
\]
This state with $|z| \ll 1$ coincides with the state $|\psi_t ;k,2
\rangle$ with $\lambda_2 t \ll 1$ (for $z = |\kappa|t$).
However, for other moments of time this simple connection is not valid
in general. For example, at the moment $\lambda_2 t = \pi/2 - 
\varepsilon$, $\varepsilon \ll 1$, the state $|\psi_t ;k,2\rangle$ of 
Eq.\ (\ref{psi_tk2}) has the form
\begin{eqnarray}
|\psi_t ;k,2\rangle & \approx & (4k+1)^{-1} \Bigl[ \sqrt{8k(2k+1)}
(1-\varepsilon^2) |0\rangle|k,2\rangle 
+ \sqrt{8k(4k+1)} \varepsilon |1\rangle|k,1\rangle \nonumber \\
& & + (1+4k\varepsilon^2) |2\rangle|k,0\rangle \Bigr] . \label{psi_eps}
\end{eqnarray}
On the other hand, the coherent state $|z;k,2\rangle$ with $|z| \gg 1$
can be written as
\begin{equation}
|z;k,2\rangle \approx (1-g^2) |0\rangle|k,2\rangle +
\sqrt{2} g |1\rangle|k,1\rangle 
+ \sqrt{(2k+1)/(2k)} g^2 |2\rangle|k,0\rangle , \label{z_g}
\end{equation}
where we have assumed $z$ to be real and positive, and 
$g = 1/(\sqrt{2k+1}z)$.
The state $|z;k,2\rangle$ of Eq.\ (\ref{z_g}) does not coincide in
general with the state $|\psi_t ;k,2\rangle$ of Eq.\ (\ref{psi_eps}).
It is seen that these two states coincide only in the limit 
$k \gg \varepsilon^{-2} \gg 1$ (with $g = \varepsilon$).

In order to understand why the system does not evolve in general 
into the coherent state $|z;k,2\rangle$ with $|z| \gg 1$, consider the 
result (\ref{nlim}) for the mean photon number 
$\langle \hat{N}_a \rangle$. Indeed, for $|z| \gg 1$, 
$\langle \hat{N}_a \rangle$ is very small, i.e., almost all the photons 
are in the signal-idler system. This result contradicts to a 
fundamental limit on the energy transfer, that exists for the evolution
of the initial state $|L\rangle|k,0\rangle$. It was first shown by 
Drobn\'{y} and Bu\v{z}ek \cite{DB94} that such a fundamental limit on 
the energy transfer exists in the multiphoton down-conversion, 
when the signal mode $\hat{b}$ is initially in the vacuum state and the 
pump mode $\hat{a}$ is excited. A particular case is the two-photon
down-conversion (i.e., the degenerate parametric amplification),
described by the Hamiltonian (\ref{h2}) with the two-photon SU(1,1)
realization (\ref{k2}). In this case only less than $3/4$ of the pump 
photons can be transferred in average from the pump to the signal mode 
\cite{DB94}.

A similar limit holds also for the evolution induced by the general 
Hamiltonian (\ref{h2}) for the initial state $|L\rangle|k,0\rangle$.
For $L=1$, we find
\begin{equation}
\langle \hat{N}_a \rangle (t) = \cos^2 (\lambda_1 t) .
\end{equation}
In this case the single photon being initially in the pump mode can 
be completely transferred to the signal mode. Therefore, there is no
prohibition on the production of the coherent states with arbitrarily
large values of $|z|$. However, for $L>1$ there
does exist the limit on the energy transfer. For $L=2$, we obtain
\begin{equation}
\langle \hat{N}_a \rangle (t) = 2 - \frac{8k \sin^2 
(\lambda_2 t)}{4k+1} - \frac{8k \sin^4 (\lambda_2 t)}{(4k+1)^2}   .
\end{equation}
The efficiency $\xi$ of the energy transfer is defined as the maximum 
mean number of photons which can be in principle transferred from the 
pump to the signal-idler system, divided by the initial number of the 
pump photons. For $L=2$, the efficiency is
\begin{equation}
\xi = 1 - \frac{1}{ (4k+1)^2 } .
\end{equation}
In particular, in the degenerate parametric amplification we have
$k=1/4$ and $\xi = 3/4$, in accordance with the results of Drobn\'{y} 
and Bu\v{z}ek \cite{DB94}. However, in the non-degenerate parametric 
amplification the efficiency $\xi$ rapidly approaches 1 as $k$ 
increases. Therefore, the coherent states with large values of $|z|$
can be produced in the limit $k \gg 1$, as we have seen from the
comparison of Eqs.\ (\ref{psi_eps}) and (\ref{z_g}).

We also find an heuristic relation between the coherent states and
the long-time evolution induced by the trilinear Hamiltonian.
Consider the state $|\psi_t ;k,2\rangle$ at the moment 
$\lambda_2 t = \pi/2$:
\begin{equation}
|\psi_t ;k,2\rangle = \frac{ |2\rangle|k,0\rangle + \sqrt{8k(2k+1)}
|0\rangle|k,2\rangle }{ 4k+1 } .  \label{eq_last}
\end{equation}
We compare this state with an even superposition of two coherent 
states, 
\begin{equation}
|z;k,2\rangle_{e} = {\cal A} \bigl( |z;k,2\rangle + |-z;k,2\rangle 
\bigr) = \frac{ |2\rangle|k,0\rangle + \sqrt{2k(2k+1)} z^2
|0\rangle|k,2\rangle }{ \sqrt{1+2k(2k+1)z^4} } ,
\end{equation}
where ${\cal A}$ is a normalization factor and we assume $z$ to be real.
It is easy to see that the superposition state $|z;k,2\rangle_{e}$
coincides with the state $|\psi_t ;k,2\rangle$ of Eq.\ (\ref{eq_last}) 
for $z^2 = 2$. In this special case the long-time dynamics can be 
described by using a discrete superposition of the coherent states.
However, any quantum state obtained during the time evolution of the
system can be expanded in the coherent-state basis by using the identity
resolution (\ref{id}).

\section{Conclusions}

In this paper we have introduced the coherent states for quantum 
systems, whose dynamics is governed by a trilinear boson Hamiltonian. 
This dynamics is complicated because the operators which compose such 
a Hamiltonian do not form a finite-dimensional Lie algebra. Therefore,
the standard formalism of Perelomov cannot be used in this case. 
Nevertheless, we have shown that in each invariant subspace there does
exist an overcomplete set of the coherent states. During the first
instants of the time evolution the system prepared in a special initial 
state will evolve into the coherent state.
Using the coherent-state basis, we have constructed the resolution of 
the identity and the analytic representation in the complex plane.
This analytic representation coincides with a double representation
based on the Glauber coherent states of the pump and on the SU(1,1)
Perelomov coherent states of the signal-idler system. 
We have studied the entanglement between the field modes and photon 
statistics of the coherent states. We have also considered connections 
between the coherent states and the long-time evolution induced by the 
trilinear boson Hamiltonian.

\acknowledgements

The author is grateful to Profs. A. Mann and J. Katriel for helpful 
discussions and thanks the referees for valuable comments.
The financial help from the Technion is gratefully acknowledged.

\vspace*{2cm}
\section*{Figure Captions}
\begin{figure}
\caption{The purity parameter $S^{{\rm pur}}_a$ of Eq.\ 
(\protect\ref{pur}) as a functions of $|z|=1/|y|$ for the coherent 
states $|y;k,L\rangle$ with $k=1$ and $L=1,3,10$. The loss of the
correlation for $|z| \gg 1$ occurs because the pump mode tends in 
this limit toward the vacuum and is then more pure.}
\label{fig:pur}
\end{figure}
\begin{figure}
\caption{The photon-number distribution $P_{a}(n)$ of Eq.\ 
(\protect\ref{Pn}) for the coherent states $|y;k,L\rangle$ 
with $k=1$, $L=10$, and various values of $|z|=1/|y|$.}
\label{fig:pnd}
\end{figure}
\begin{figure}
\caption{The mean photon number $\langle \hat{N}_a \rangle$ of Eq.\ 
(\protect\ref{Nm}) [curve (a)], and the variance 
$\langle (\Delta \hat{N}_a)^2 \rangle$ of Eq.\ (\protect\ref{Nv}) 
[curve (b)] as functions of $|z|=1/|y|$ for 
the coherent state $|y;k,L\rangle$ with $k=1$ and $L=10$. As 
$|z|$ increases, the pump mode is depleted and approaches
the vacuum in the limit $|z| \rightarrow \infty$.}
\label{fig:nn}
\end{figure}

\end{document}